\documentclass[11pt,twoside]{article}
\usepackage{asp2010}

\resetcounters

\begin{document}

\title{External Use of TOPCAT's Plotting Library}
\author{M.~B.~Taylor}
\affil{H.~H.~Wills Physics Laboratory, University of Bristol, U.K.}

\begin{abstract}
The table analysis application TOPCAT uses a custom
Java plotting library for highly configurable
high-performance interactive or exported visualisations
in two and three dimensions.
We present here a variety of ways for end users or application developers
to make use of this library outside of the TOPCAT application:
via the command-line suite STILTS or its Jython variant JyStilts,
via a traditional Java API, or by programmatically assigning
values to a set of parameters in java code or using some form of
inter-process communication.
The library has been built with large datasets in mind;
interactive plots scale well up to several million points,
and static output to standard graphics formats is possible for
unlimited sized input data.
\end{abstract}

\section{Introduction}

TOPCAT\footnote{\url{http://www.starlink.ac.uk/topcat/}} \citep{topcat}
is a desktop GUI application for analysis of
tabular data, particularly source catalogues.
Among other capabilities it provides high-performance interactive
visualisation for large (and small) datasets.
The plotting capabilities are focussed on representations of
point clouds in two or three dimensions, with special attention
to large (many row) and high-dimensional (many column) datasets.
Many configuration options are offered.

The visualisation is supported by a custom Java plotting library,
written from scratch for TOPCAT v4.0 \citep{2014ASPC..485..257T}.
The recent release (v3.0) of the
STILTS\footnote{\url{http://www.starlink.ac.uk/stilts/}}
command-line suite
exposes all of TOPCAT's visualisation capabilities
in ways that can be harnessed from outside of the application
itself.

\section{Features}

Each plot is composed of a {\em plot surface\/}, defining the geometry
and decoration of the axes and zero or more {\em plot layers\/}.
Different layers may use different data sets, allowing all kinds
of overplotting.
Currently defined plot surface types are
2d Cartesian,
Celestial (offering a number of projections),
3d Cartesian,
Spherical polar and
2d with a time axis.
Many layer types are available including
   scatter plot,
   lines,
   contours,
   analytic function,
   error bars,
   ellipses,
   pair links,
   text labels,
   vectors,
   sized markers,
   histogram,
   spectrogram,
   colour coding by density or additional coordinates,
and more.
Each surface and layer type offers a large number of configuration options.
The structure of the package allows easy extensibility,
so additional surface types, layer types and configuration options
are expected in future releases or can be plugged in at runtime.

The library has been written with performance and scalability as
a primary aim.  Interactive navigation typically works well up to
a few million points.  When generating static output files,
datasets of unlimited size can be used without large memory requirements.
An all-sky density plot of 2 billion rows takes of the order of
30 minutes to produce.
Moreover options such as hybrid scatter/density plots are available
to provide visually meaningful graphics from very large datasets.

\section{Implementation}

The STILTS package \citep{2006ASPC..351..666T}
is a suite of command-line tools built on the same
libraries as TOPCAT, aiming to provide access to the same capabilities.
While providing alternative interfaces to the same data access, analysis,
cross-matching, and I/O facilities is a straightforward enough goal,
in fact a large proportion of the complexity of both the command-line
and GUI tools is devoted to the user interface itself,
so providing the same functions from both contexts is not trivial.

The problem is particularly challenging in the case of the new
visualisation tasks.
Although of the order of a hundred plotting parameters and configuration
options are available to allow the construction of complex and finely
tuned plots, a simple plot with default settings can be specified
with a very small number of parameters ($\sim 4$).
Managing this complexity in both the implementation and the
user interface requires careful design and a significant amount of logic.
Some notable problems include
defining command-line-friendly (i.e.\ string) representations
for all parameter names and values,
providing both interactive and reference
documentation to the user for all the numerous options,
and allowing easy extensibility (new or modified plot types and
configuration options) without requiring changes to hand-crafted
UI or graphical output code.

For this reason the plotting infrastructure is built on a pluggable
and object oriented set of interfaces that allow the host application
to construct a graphical or text user interface, as well as its
user documentation, by interrogating the plotting objects themselves,
with little detailed knowledge of the plotting capabilities.

\begin{figure}
% \rule{\textwidth}{0.5mm}
\begin{tabular}{lclc}
\makebox[0cm]{\em a)} &  & \makebox[0cm]{\em b)} & \\
%
% a) evanthia
   & \hspace*{-5mm}
     \includegraphics[width=0.45\textwidth]{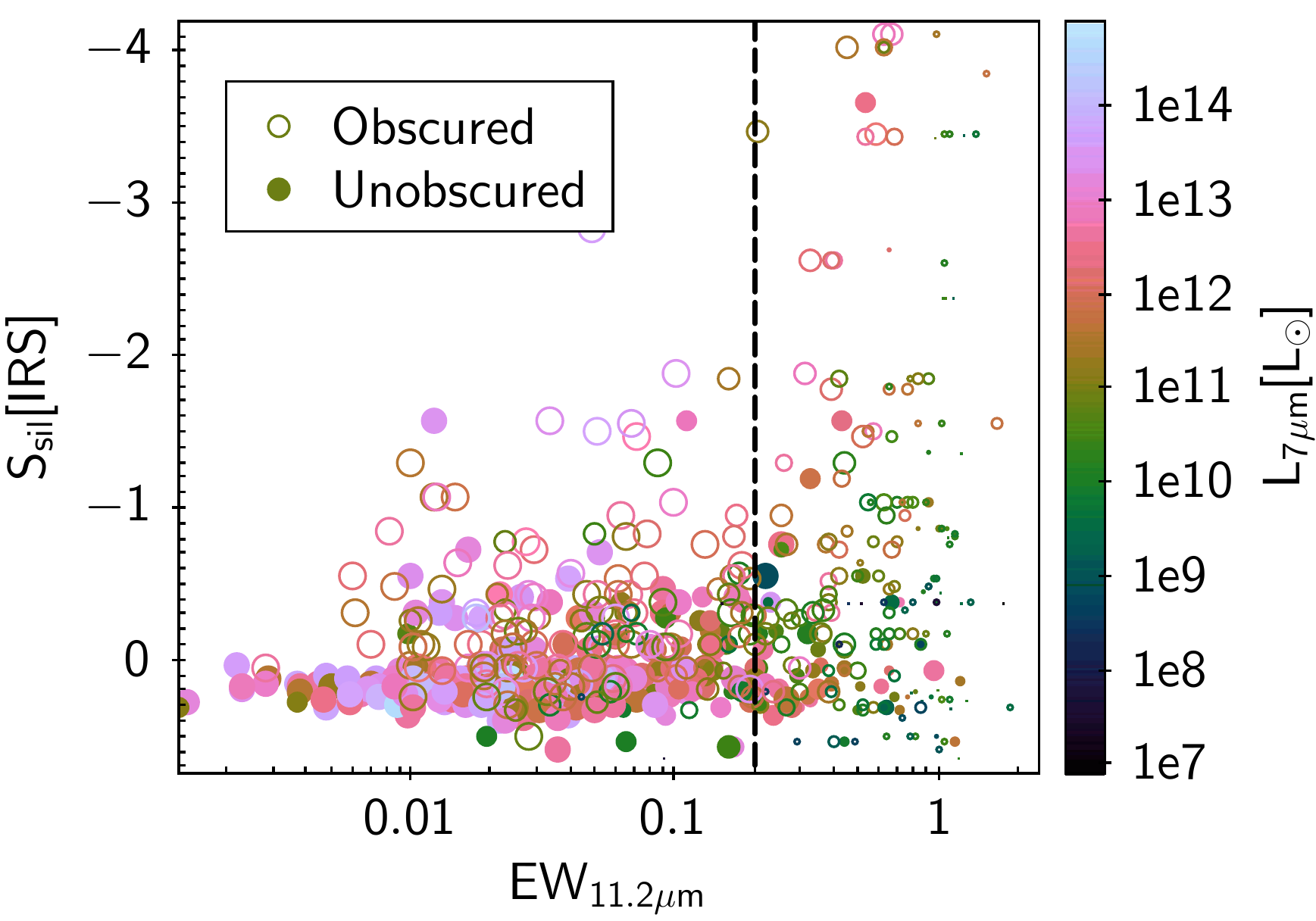}  &
%
% b) mw
   & \hspace*{-5mm}
     \raisebox{9mm}{\includegraphics[width=0.5\textwidth]{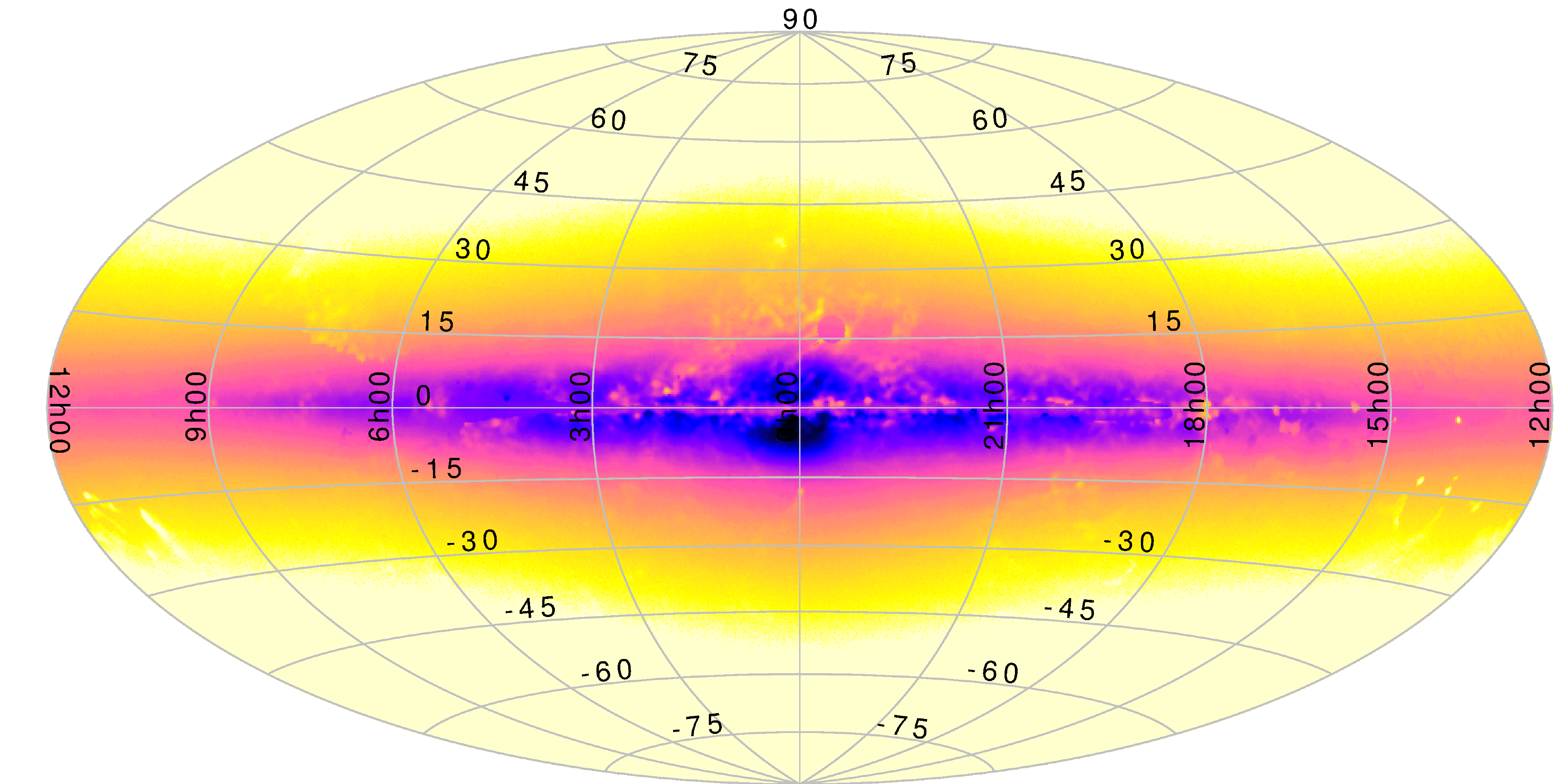}} \\[5mm]
   \makebox[0cm]{\em c)} &  & \makebox[0cm]{\em d)} & \\
%
% c) vectors
   & \hspace*{-5mm}
     \includegraphics[width=0.4\textwidth]{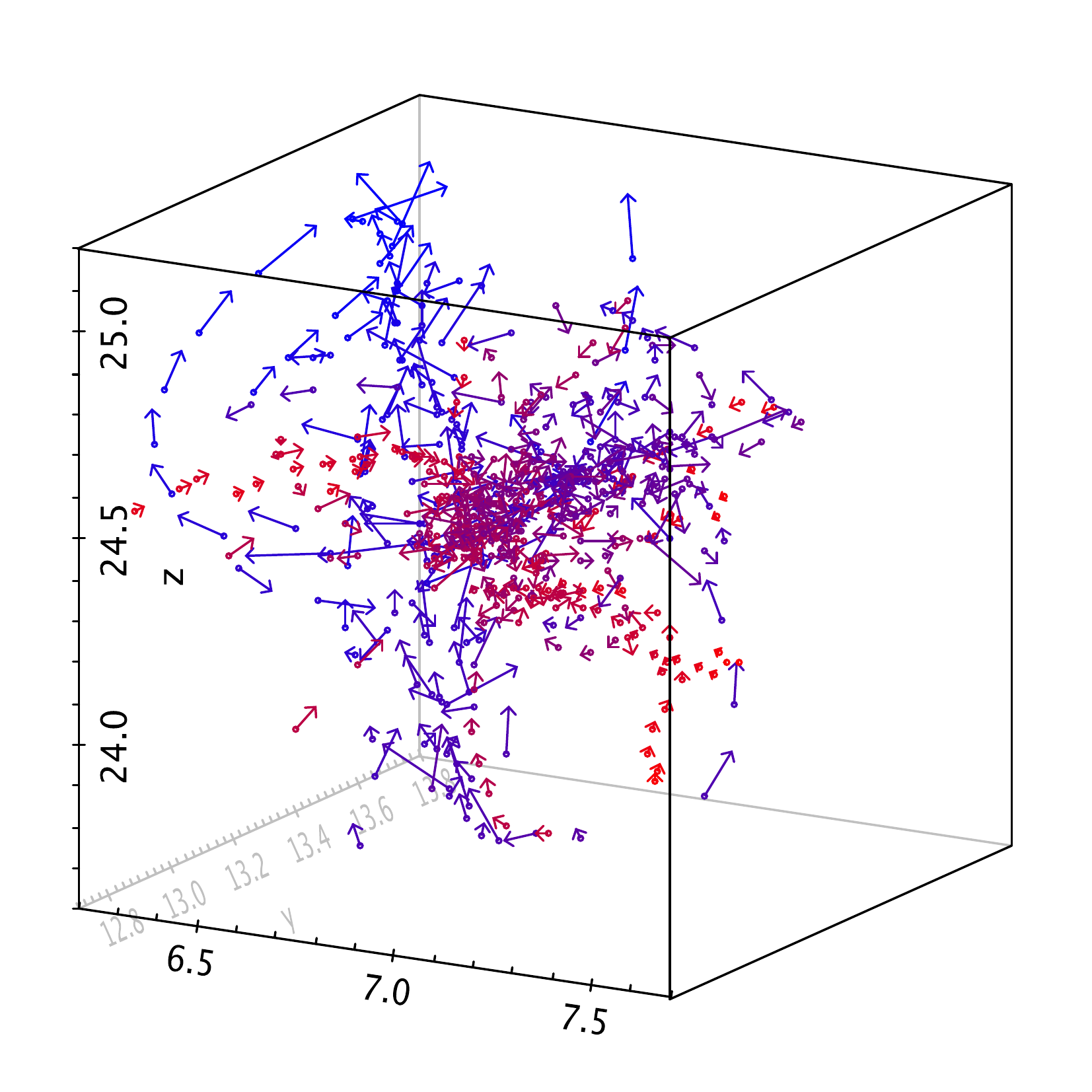} &
%
% d) TimeStack
   & \hspace*{-5mm}
     \raisebox{8mm}{\includegraphics[width=0.45\textwidth]{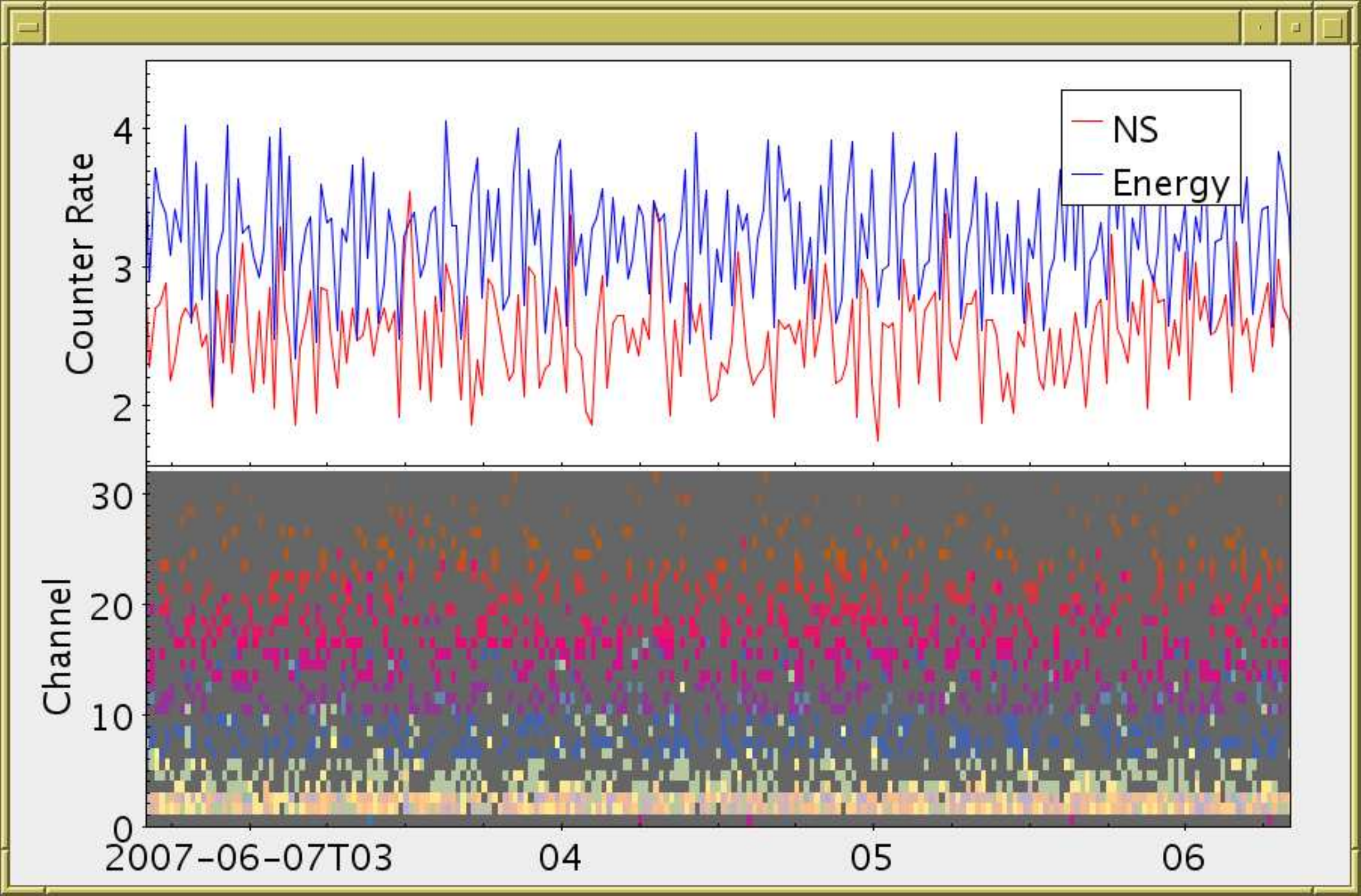}} \\[5mm]
   \makebox[0cm]{\em e)} &  & \makebox[0cm]{\em f)} & \\
%
% e) paul1
   & \hspace*{-5mm}
     \includegraphics[width=0.4\textwidth]{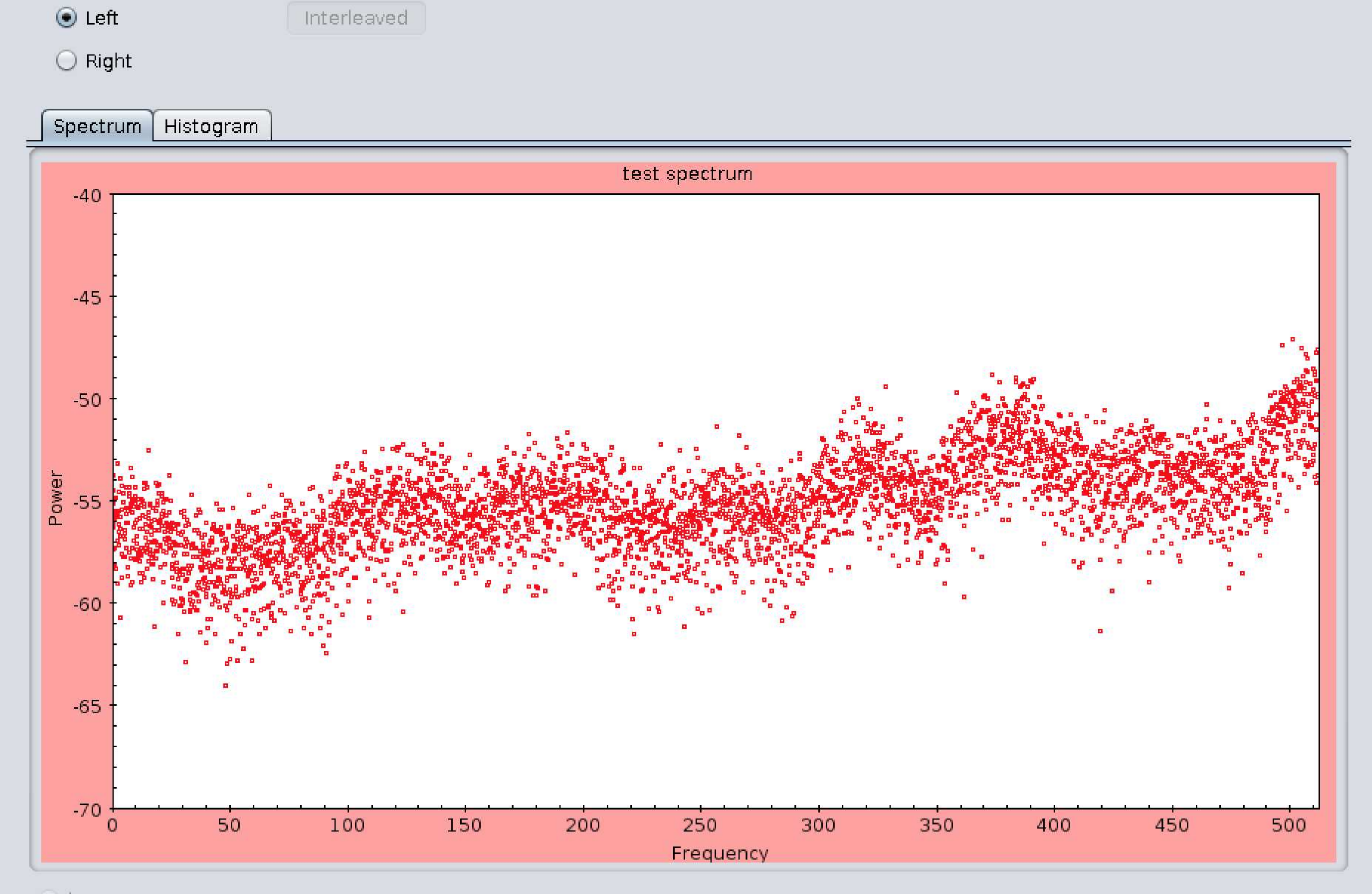} &
   & \hspace*{-5mm}
%
% f) smilgys
     \includegraphics[width=0.5\textwidth]{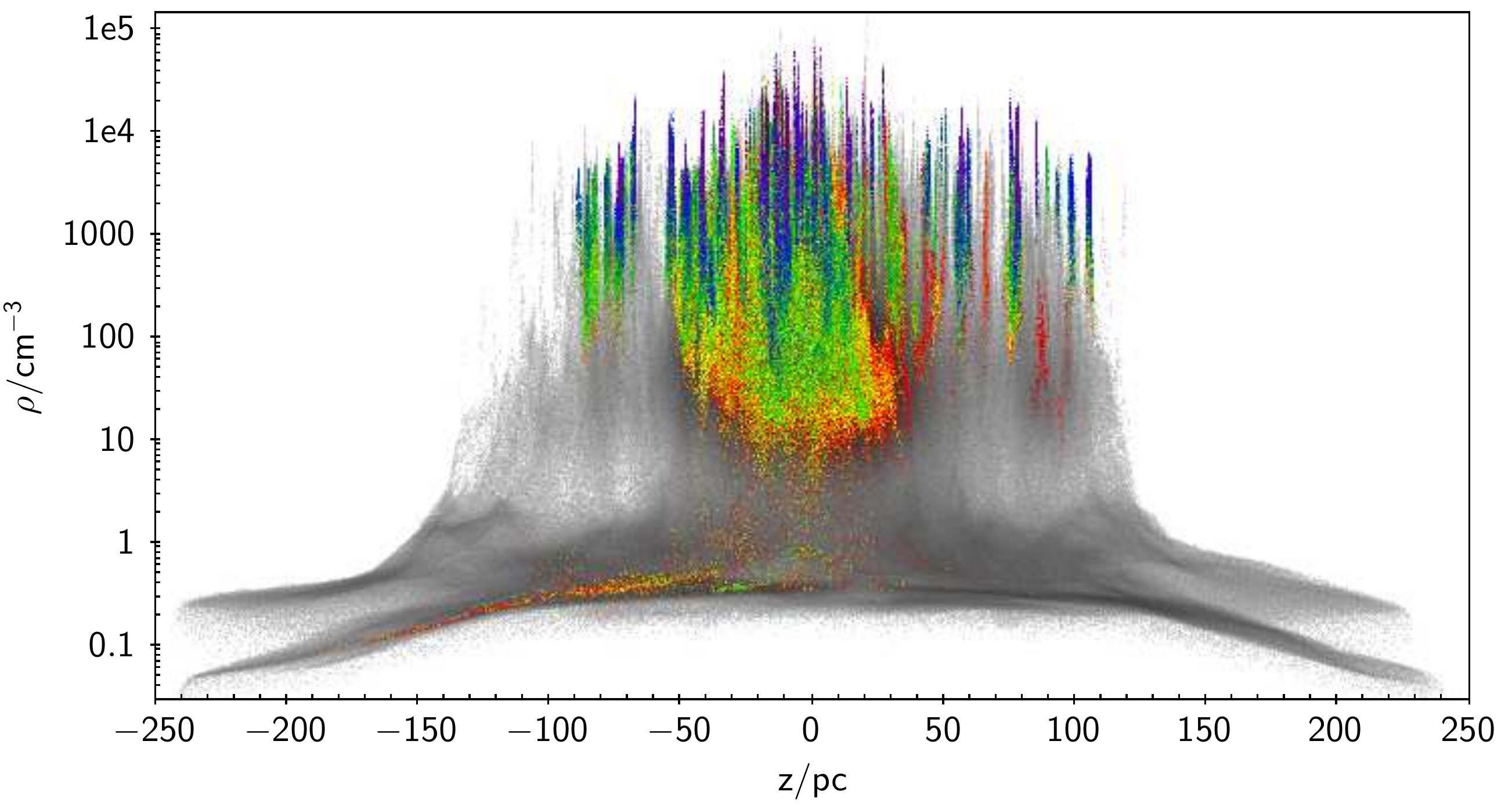}
\end{tabular}
\caption{Example plots.
(a) Points coded by marker shape, size and colour (E.~Hatziminaoglou, ESO).
(b) Density map for the GUMS-10 milky way simulation dataset;
    2 billion rows plotted in around 30 minutes.
(c) Simulation data, with points and velocities represented.
(d) Spectrogram and data samples plotted on an interactively scrollable
    time-axis.
(e) Real-time display of an observed spectral data, 8 thousand points
    refreshed, easily, at 1\,Hz (P.~Harrison, Jodrell Bank).
(f) SPH simulation data, 14 million points, plotted in 8 seconds
    (R.~Smilgys, St.~Andrews)
}
\end{figure}

\section{Invocation}

A number of options are available for invoking the plotting capabilities
from outside the TOPCAT application.

First, the STILTS command-line interface allows generation of plots
from a command line such as a unix shell, for instance:
{
\small
\begin{verbatim}
   stilts plot2plane layer_1=mark in_1=data.fits x_1=BMAG-RMAG y_1=BMAG
\end{verbatim}
}
\noindent
will generate a simple scatter plot based on the given arithmetic expressions
involving column names from the given input table (a FITS file in this case).
The JyStilts package provides a Jython variant of this interface
to allow use with python syntax and control structures.

Java applications may make analogous invocations by specifying
the parameter name/value pairs as entries in a supplied hash
(a java {\tt Map} object).  In this case each parameter value may be
supplied either as a string, using the same rules as on the command-line,
or as an object corresponding to the documented parameter type.
For instance the value of a parameter supplying input table data
may be either the name of a file containing a table in one of
the supported formats (FITS, VOTable, CSV etc)
or a programmatically constructed object implementing the
{\tt StarTable} interface, which could be a custom iterable over
values in an existing memory-resident data structure.

Finally, java application code may use the low-level API documented
in javadocs to specify all the details of the plot explicitly.
Since this does not benefit from the extensive defaulting
offered by the variants of the key/value interface it is quite
verbose (50--100 lines to produce the example plot above)
but it does provide detailed control and compile-time checking.

\section{Graphics Output}

Once a plot has been specified to the library it can be output
in two basic ways.

First, it can generate a live interactive plot.
From the command line, this appears in the form of a window popped
up on the screen, and from the API as a Swing {\tt JComponent} widget
that can be incorporated into the host application's GUI.
In either case this window can be resized at will and
offers the user sophisticated interactive navigation capabilities
(pan, zoom and for 3d plots rotation and recentering).

Alternatively, it can generate a static output file in one of
several supported bitmap (PNG, GIF, JPEG)
or vector (PDF or PostScript) graphics formats.
Options for convenient and efficient production of animations
% (sequences of related static images)
are also provided.

\section{Future Work}

The STILTS plotting library is still somewhat experimental,
and there may be adjustments to the API in future releases
to accomodate more surface and layer types and configuration options.
Full reference and some tutorial documentation is already available
in the STILTS
user document, but improved tutorial documentation and more code
examples will be forthcoming.  Efficiency improvements will be
explored by use of multithreading.  Integration with the TOPCAT
application will be improved by showing the command-line settings
corresponding to plots set up in the GUI.

We hope that the library presented here will be of use for scientists
and application developers with requirements for plotting 
from java-friendly contexts.

\bibliography{P6-5}

\end{document}